# Photo-driven Molecular Wankel Engine, $B_{13}^+$

Jin Zhang[†,*], Alina P. Sergeeva[‡], Manuel Sparta[†], Anastassia N. Alexandrova[†,*]

[†] Department of Chemistry and Biochemistry, University of California Los Angeles, 607 Charles E. Young Drive East, Box 951569, Los Angeles, CA 90095-1569, USA

[‡] Department of Chemistry and Biochemistry, Utah State University, 0300 Old Main Hill, Logan, Utah 84322-0300, USA

Supporting Information Placeholder

**ABSTRACT:** We report a molecular Wankel motor, the dual-ring structure $B_{13}^+$, driven by circularly-polarized infrared electromagnetic radiation, under which a guided uni-directional rotation of the outer ring is achieved with rotational frequency of the order of 300 MHz.

Synthetic molecular motors, capable of delivering controlled movement upon energy input, is one of the key building blocks in nano-machinery.[1] The major energy sources of molecular motors are from chemical reactions,[1(a)] photon beams[1(b)] or electric current,[1(f)] which are converted to mechanical forces through the excitation of the electronic states of the molecule. The energy scale of the electronic excitation is normally two orders of magnitude larger than the molecular vibrational frequencies. To reduce the heat dissipation and increase the energy utilization efficiency, a motor running purely on the electronic ground state (GS) potential energy surfaces is highly desirable.

Being neither molecules nor pieces of solids, nano-clusters are unique chemical species with remarkable novel structures, chemical bonding, and unexpected reactivity.[2] Recently, two planar boron clusters, $B_{19}^-$ and $B_{13}^+$, have been demonstrated to undergo a rare fluxional behavior: the outer ring rotates almost freely with respect to the inner ring.[3-6] These clusters are anomalously stable species of low reactivity,[4,7] similar to their prototypical organic aromatic congeners: benzene and coronene.[2b,2c,4,6,9] The low energy barrier that made the in-plane rotation possible was due to the delocalized bonding, which rendered concentric double aromaticity of both equilibrium and transition states of the clusters.[5,6] This peculiar fluxionality of $B_{19}^-$ and $B_{13}^+$ at room temperature has gained them the name of "molecular Wankel motors".[5,6]

In equilibrium, driven by thermal fluctuations, the outer ring rotation is bi-directional, and has a time scale of the order of tens of pico-seconds at 300 K. To utilize $B_{13}^+$ as an engine for nano-machinery, one needs an energy source to drive the molecule out of equilibrium. In this Communication, we show that a uni-directional rotation of the outer ring of $B_{13}^+$ can be achieved with a circularly-polarized infrared (IR) laser, rendering a photo-driven molecular Wankel motor running on the electronic GS potential energy surfaces with a rotational period of a few pico-seconds.

The ground state (GS) geometry of $B_{13}^+$, which exhibits $C_{2v}$ symmetry, is shown in Figure 1. In equilibrium, $B_{13}^+$ has a small intrinsic dipole moment (0.4 Debye) along the $C_{2v}$ axis, through which we define the square ring as the head of the molecule. In equilibrium, a complete 360° rotation of the outer ring is decomposed into 30 identical elementary moves (see Figure 3 of Ref. 6 for detailed description).

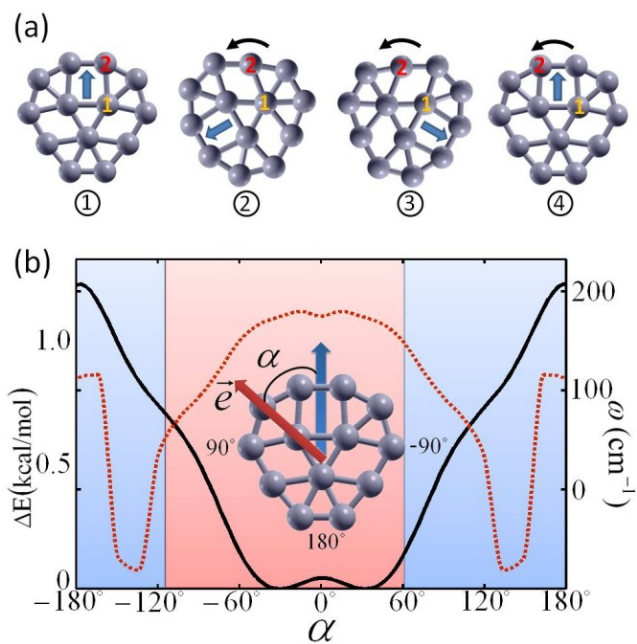

Figure 1. (a) Schematic of the outer ring rotation. Three identical counterclockwise elementary rotations are involved in the sequence 1-2-3-4. The orientation of the molecule is defined by the blue arrow, which is aligned with the $C_{2v}$ axis of the molecule. (b) Relative ground state energies (left axis, solid black line) and vibrational frequencies (right axis, red dashed line) of the elementary rotation of the $B_{13}^+$ as a function of the angle α when a constant electric field ($|\vec{e}| = 2.6 \times 10^9$ V/m) is applied. α (<180°) is defined as the angle between the field direction and the molecule. The blue (red) region indicates that a counterclockwise rotation depicted in (a) is favored (unfavored) at a given angle α (see text).

Each elementary move induces a slight relative movement between the inner ring and the outer one. In Figure 1 (a), through the sequence 1-2-3-4, all atoms in the outer ring $B_{10}$ are rotated counterclockwise by one atomic position. A 36°-rotation is thus completed through three consecutive elementary moves. During each move, a re-orientation of the molecule takes place and rotates the head counterclockwise by 120°. This discretized re-orientation of the molecule is critical to the guided uni-directional rotation of the outer ring, as will be shown below.

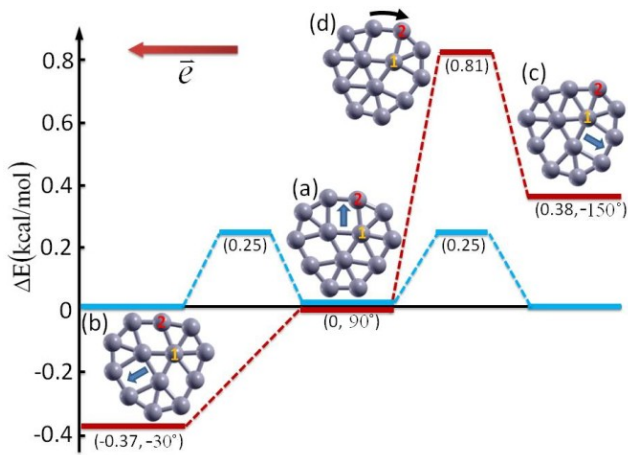

Figure 2. Energy profile of two elementary rotational moves (clockwise and counterclockwise) of the $B_{13}^+$ outer ring. (a) the initial geometry; (b,c) the final geometry after an elementary (counter-)clockwise move; (d) the transition state for the clockwise rotation. The blue line shows the transition path without electric field. When a constant electric field ($|\vec{e}| = 2.6 \times 10^9$ V/m) is applied, the counterclockwise rotation (a→b) is strongly favored over the clockwise one (a→c) with the transition path shown in the red line. There is almost no barrier for the counterclockwise move. The relative energies and angle α are given in parentheses. The marked B atoms and blue arrows are guides for eyes to illustrate the outer ring rotation and changes in the orientation of the molecule.

When subjected to an electric field, the molecule's delocalized orbitals are easily polarized. We found that the molecule has a tendency to line-up with the field to minimize the field component normal to the molecular plane. In the following we limit our discussion to the 2D case where the electric field and the molecule are coplanar. We characterize the orientation of the molecule with respect to the field direction by angle α (see Figure 1 (b)). By studying the GS energies as a function of α, we found two energy minima which approximately correspond to α=±30°. Because the energy scale of the thermal fluctuation at room temperature is 0.6 kcal/mol, a fairly large portion of the bottom region with relatively small α is thermally reachable. For larger α, the configuration becomes gradually energetically unfavorable.

There are two ways to lower the total energy in the regions of large α: 1) rotating the molecule as a whole until α falls into the favorable region; 2) rotating the outer ring with an elementary rotation and change the molecule's orientation by ±120°. Which route is more likely to occur? Using vibrational analysis, we found that the harmonic frequency of the outer ring rotation in thermal equilibrium is 135 cm$^{-1}$. On the other hand, at room temperature, according to the equi-partition of the kinetic energy, the frequency of the rigid rotation corresponds to about 21 cm$^{-1}$. The separation of the two energies suggests that it is possible to rotate the molecule through the second route if the frequency of the driving force is close to 135 cm$^{-1}$.

Route 1 and route 2 do not only differ in energies but also in rotation characteristics. In route 1, the molecule rotates as a whole continuously, descending along the energy curve until reaching one of the minima in Figure 1(b). In route 2, the outer ring rotates with respect to the inner triangle through elementary moves, each of which rotates the molecule's head by a fixed amount ±120° (see Figure 1(a)). There is no guarantee that the final state has a lower energy. For example, for any $|\alpha_0| < 60°$, an elementary rotation towards either direction gives $|\alpha_0'| \equiv |\alpha_0 \pm 120°| > 60°$, which suggests that any re-orientation of the molecule through route 2 starting with $\alpha_0$ always increases the total energy. Statistically, this region may be considered as a forbidden-zone for an elementary rotation. The forbidden region for the counterclockwise move depicted in Figure 1(a) can be similarly deduced. In Figure 1(b), the energetically allowed region is shown in blue whereas the unfavored region is in red. In each elementary rotation, the work/heat produced can be estimated from the difference in the GS energies before and after the rotation, which reads approximately 0.5 kcal/mol. The heat can be quickly dissipated through IR radiation to a thermal bath at 300 K (0.6 kcal/mol) without causing significant increase in the molecular temperature.

When a constant electric field is applied, the moderately expressed dipole moment of $B_{13}^+$ causes unevenly populated states as a function of the molecule's orientation. This uneven population paves the way to realize a uni-directional rotation of the molecule. Moreover, under an electric field the frequency of the elementary rotation varies drastically between −80 cm$^{-1}$ and 160 cm$^{-1}$ (see Figure 1(b)). The negative frequency indicates an unstable geometry which deforms (rotates) spontaneously. Here we consider the transition path of two elementary rotations towards opposite directions with an initial state of α=90°, shown in Figure 2. For other initial states with positive frequencies and $|\alpha_0| > 60°$, similar conclusions were found.

Without electric field, the clockwise and counterclockwise rotations are equally possible with a small barrier of 0.25 kcal/mol separating the initial and the final state. However, once a constant electric field is applied, the final states for clockwise/counterclockwise rotation differ in energy significantly. What's more interesting is the fact that the transition state barrier for the counterclockwise rotation almost completely disappears, whereas the barrier for clockwise rotation increases more than three-fold, to 0.81 kcal/mol. This transition path is ideal for rotating the outer ring uni-directionally.

The final step for a molecular Wankel motor design is to find a closed loop energy path along which the molecule is driven to work repeatedly. Here we propose to drive the molecule with a circularly-polarized IR radiation along the path illustrated in Figure 3. The circularly-polarized electromagnetic wave is approximated as a (counter)-clockwise rotating electric field at fixed frequency with constant modulus. In Figure 3(a), the ini-tial state is one of the states in the forbidden region, with some angle $\alpha_1$ ($|\alpha_1| < 60°$). As the electric field rotating counterclockwise, $\alpha_1$ gradually increases to some point $\alpha_2$, at which the total energy is too high to maintain the molecule's geometry stable. This phase P1 does not involve molecular re-orientation (elementary rotation). Only the total energy of the molecule changes with the rotating field. The phase P2 is the re-orientation process at $\alpha_2$. Compared to the P1 phase, the time elapse of P2 is much shorter. Once the elementary rotation is finished, a new angle $\alpha' = \alpha - 120°$ is introduced. As long as $\alpha_2 < 180°$, the new angle $|\alpha'| < 60°$. The system is restored to the P1 phase, and the loop is closed.

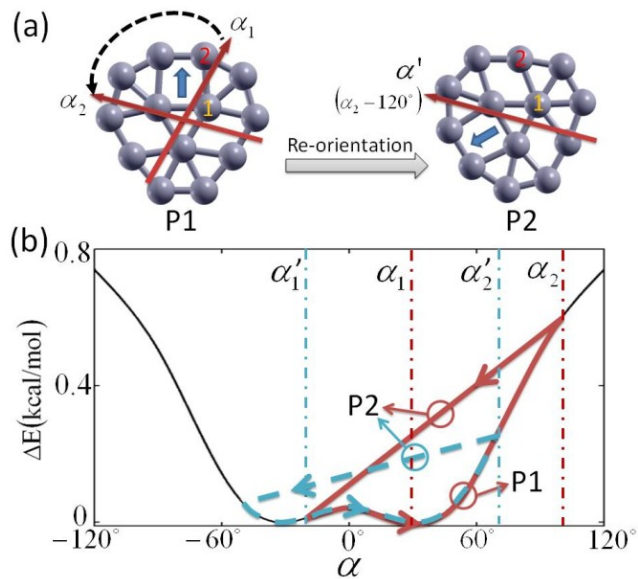

Figure 3. The elementary rotational move of the outer ring when guided by a circularly-polarized electric field. (a) In phase P1, the field starts at $\alpha = \alpha_1$ ($|\alpha_1| < 60°$) and rotates counterclockwise. There is no re-orientation of the molecule occurring in P1. Once $\alpha$ reaches above $60°$, the molecule's tendency of doing a counterclockwise elementary rotation increases. At some point $\alpha_2$ ($60° < \alpha_2 < 180°$), the molecule re-orientates itself through an elementary rotation to lower its total energy. The new phase P2 has an angle $\alpha' = \alpha_2 - 120°$ owing to the re-orientation, which brings $|\alpha'| < 60°$. The cycle is completed and the system is back to P1 phase. (b) Two example work loops of the proposed Wankel motor: $\alpha_1 \to \alpha_2 \to \alpha'_1 \to \alpha'_2$. Marked with vertical dash-dotted lines, the angles $\alpha_{1,2}$ ($\alpha'_{1,2}$) are the starting points of phase P1/P2 on the red solid line (blue dashed line). In both loops, the lower branches are in P1 phase while the upper ones are in P2.

In Figure 3(b), an example of the above proposed work loop is shown. The system starts from $\alpha_1$ and is driven by a counterclockwise rotating field along the red line. At some point $\alpha_2$, which is either randomly chosen by stochastic processes or due to the negative frequencies (unstable structure), the outer ring undergoes a counterclockwise elementary rotation and re-orientates the molecule with a new angle $\alpha'_1 = \alpha_2 - 120°$. The rotating field then keeps driving the system along the blue dashed line. And again at point $\alpha'_2$, the outer ring performs another counterclockwise rotation. Over time, the outer ring continues to rotate counterclockwise relative to the inner ring through a series of elementary rotations induced by closed loops similar to the red and blue lines shown in Figure 3(b).

From above discussion, one sees that the (counter-)clockwise rotating field promotes a (counter-)clockwise rotation of the $B_{13}^+$ outer ring with respect to the inner one. What happens if the initial state falls out of the proposed phase, e.g. $-180° < \alpha_1 < -60°$ in Figure 3? The first possibility is that the molecule performs an elementary rotation since it may not be stable or its total energy is high. In either direction, it will fall into one of the proposed phases. If the re-orientation does not happen, the molecule remains largely structurally unchanged. As the field rotates, $\alpha_1$ keeps increasing until it reaches the P1 phase. In either case, the molecule's outer ring rotation is synchronized.

To verify the proposed working principle of the molecular Wankel motor, we performed a Born-Oppenheimer molecular dynamics (BOMD) of a single $B_{13}^+$ ring at 600 K. The rotating electric field is simulated by rotating the molecule rigidly under a constant electric field at 3.1 THz (approx. 100 cm$^{-1}$). In principle, an elementary rotation requires the electric field to rotate by $120°$. In other words, a $360°$ rotation of the outer ring needs at least 10 loops of electric field rotation. In practice, because of thermal fluctuations, this gear-ratio is expected to be higher than 10:1. With above parameters, we found the ratio was about 13:1. The video for the guided rotation of the outer ring can be found in the supplementary material where a total of 11 electric field rotations were performed. Finally, to avoid exciting other degrees of freedom, the applicable frequency range of the IR radiation for the Wankel motor should be restricted to between 60 cm$^{-1}$ and 160 cm$^{-1}$. The electric field strength we used in the simulation can be achieved via ultraintense lasers experimentally.

In summary, we propose a model molecular Wankel motor $B_{13}^+$ driven by circularly polarized IR radiation near 3 THz, as a potential building block for nano machines. The rotating frequency of the outer ring of is approximately 250 MHz. A similar principle could be used for creating other uni-directional molecular motors, based on pure boron, and other types of clusters.

**Computational Details** The calculations were performed at density functional theory level with the Perdew-Burke-Ernzerhof (PBE) functional,[10] plane-wave basis set and ultrasoft pseudopotentials implemented in *Quantum Espresso*.[11] For each B atom, 2s- and 2p-states occupied by three electrons were treated as valence states. The cutoff energy for the plane-wave basis set is 28 Ry. To minimize the long-range interaction among the periodic images, the Martyna-Tuckerman method[12] was adopted with an inter-molecular distance of 15 Å. The electric field was applied as a sawtooth external potential with field strength fixed at $|\vec{e}| = 2.6 \times 10^9$ V/m in the molecular region. BOMD was used to study the rotation of the molecule at finite temperatures (300-600 K). The rigid-body translational and rotational degrees of freedom were frozen by setting the initial velocities in the BOMD simulations appropriately while the Boltzmann energy distribution was retained.

## ASSOCIATED CONTENT

**Supporting Information**. This material is available free of charge via the Internet at http://pubs.acs.org.

## AUTHOR INFORMATION

### Corresponding Authors
zhang.jin@chem.ucla.edu, ana@chem.ucla.edu

### Author Contributions
The manuscript was written through contributions of all authors. All authors have given approval to the final version of the manuscript.

### Funding Sources
Financial support for this work was provided through the ACS PRF grant 51052-NDI6 (ANA). Alina P. Sergeeva is grateful to NSF for the support through the CHE-1057746 grant.

## ACKNOWLEDGMENT
Calculations were performed on the UCLA Hoffman2 shared cluster.

## ABBREVIATIONS
GS, ground state; IR, infrared; BOMD, Born-Oppenheimer molecular dynamics.

SYNOPSIS TOC

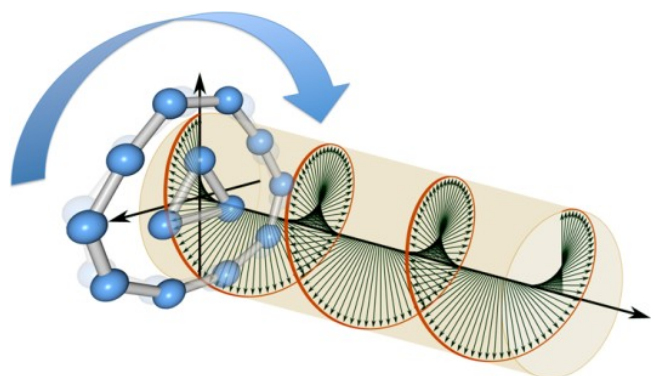